\documentclass[twocolumn,twoside,slac_two]{revtex4}
\usepackage{graphicx}
\usepackage{fancyhdr}
\begin{document}

\title{Multiwavelength Spectral Study of 3C 279 in the Internal Shock
  Scenario}

%

\author{M. Joshi, S. Jorstad, A. Marscher$^{1}$,
  M. B\"{o}ttcher$^{2}$, I. Agudo$^{1 \& 3}$, V. Larionov$^{4}$,
  M. Aller$^{5}$,\\ M. Gurwell$^{6}$, A. L\"{a}hteenm\"{a}ki$^{7}$}

\affiliation{Boston University, Boston, MA, USA$^{1}$, Ohio
  University, Athens, OH, USA$^{2}$, IAA, Granada,
  Spain$^{3}$,\\ St. Petersburg State University, St. Petersburg,
  Russia$^{4}$, University of Michigan, Ann Arbor, MI,\\ USA$^{5}$,
  SAO, Cambridge, MA, USA$^{6}$, Mets\"{a}hovi Radio Observatory,
  Kylm\"{a}l\"{a}, Finland$^{7}$}

\begin{abstract}
We have observed 3C~279 in a $\gamma$-ray flaring state in November
2008. We construct quasi-simultaneous spectral energy distributions
(SEDs) of the source for the flaring period of 2008 and during a
quiescent period in May 2010. Data have been compiled from
observations with Fermi, Swift, RXTE, the VLBA, and various
ground-based optical and radio telescopes. The objective is to
comprehend the correspondence between the flux and polarization
variations observed during these two time periods by carrying out a
detailed spectral analyses of 3C~279 in the internal shock scenario,
and gain insights into the role of intrinsic parameters and interplay
of synchrotron and inverse Compton radiation processes responsible for
the two states. As a first step, we have used a multi-slice
time-dependent leptonic jet model, in the framework of the internal
shock scenario, with radiation feedback to simulate the SED of 3C~279
observed in an optical high state in early 2006. We have used physical
jet parameters obtained from the VLBA monitoring to guide our modeling
efforts. We briefly discuss the effects of intrinsic parameters and
various radiation processes in producing the resultant SED.

\end{abstract}

\maketitle

\thispagestyle{fancy}


\section{\label{intro}Introduction}
Blazars are well known for their variability and power of polarized
radiation across a wide range of the electromagnetic spectrum
\citep[]{df2007, js2007, js2005}. In some cases, the flux can vary on
timescales as short as an hour or less \citep[see e.g.,][]{ga1996,
  at2007}. Blazars exhibit a doubly-peaked spectral energy
distribution (SED), in which the low-energy component could extend
from radio through UV or X-rays while the high-energy component
extends from X-rays to $\gamma$-rays. The low-frequency component of
the SED is almost certainly due to synchrotron emission from
nonthermal, ultra-relativistic electrons. The high-frequency
component, on the other hand, is a result of inverse Compton
scattering of seed photons by the same ultra-relativistic electrons
producing synchrotron emission (in a leptonic jet scenario). In this
case, the seed photons could be the synchrotron photons produced
within the jet (synchrotron self Compton, SSC) \citep[]{mg1985,
  gm1998}, and/or external photons entering the jet from outside (the
EC process) \citep[e.g.,][and references therein]{bm2007}. The
spectral variability patterns and SEDs are key ingredients in
determining the acceleration of particles and the time-dependent
interplay of various radiation mechanisms responsible for the observed
emission.

Another defining characteristic of blazars is the high degree of
linear polarization at optical wavelengths. Many bright $\gamma$-ray
blazars that are in the \textit{Fermi-LAT} Bright $\gamma$-Ray Source
List \citep{aa2009} have shown spectral and linear polarization
variability \citep[]{ma2010, df2009, gd2006}. Linear polarization at
mm, IR, and optical wavelengths tends to exhibit similar position
angles and sometimes correlation across these wavebands, often with
some time delay \citep[]{df2009, js2007, ls2000, gss1996,
  gs1994}. Such correspondence between the variation in polarization
and flux across a wide range of the electromagnetic spectrum, combined
with VLBI imaging, can be used to identify the location of variable
emission at all wavebands and shed light on the physical processes
responsible for the variability \citep{ma2010}.

The blazar 3C~279, located at a redshift of 0.538 \citep{br1965}, is
one of the most prominent and well-studied blazars. This is due to its
highly variable nature (change in magnitude $\Delta m \sim 5$ at
optical bands) at all wavelengths and high optical polarization up to
45.5\% observed in the U band \citep{me1990}. Intensive
multiwavelength campaigns \citep[see, e.g.,][]{lv2008, cr2008, co2007}
and theoretical efforts \citep[e.g.,][]{bp2009, bm1996} have led to
some important conclusions about the physical properties of
3C~279. Chatterjee et al. (2008) showed that the flux variability in
3C~279 has been found to be significantly correlated at X-rays,
optical R band, and 14.5 GHz wave bands, which also suggests that
nearly all X-rays are produced in the jet. The X-ray flux has also
been associated with superluminal knots, as suggested by correlation
with the flux of the core region in the 43 GHz VLBA images
\citep{cr2008}. Nevertheless, the nature and origin of its high-energy
emission and the relationship of its behavior to the physical aspects
of the jet remain elusive \citep[]{lv2008, co2010}. In addition, the
correspondence between high- and low-energy emission is also not very
well understood.

Here, we aim to understand the physical state of 3C~279 at different
flux levels and look for a correspondence between the flux and
polarization variation observed during the flaring state of November
2008 and quiescent state of May 2010. This can be achieved by carrying
out a detailed spectral analyses of 3C~279 under the internal shock
scenario, and understanding the role of intrinsic parameters and
interplay of synchrotron and inverse Compton radiation processes in
shaping the corresponding spectra of the two states.

As a first step toward comparing the physical state of 3C~279 at
different flux levels, we use the multi-slice time-dependent leptonic
jet model of Joshi \& B\"ottcher (2011) (in the framework of internal
shock scenario) with radiation feedback to simulate the SED of 3C~279
corresponding to the optical high state of early 2006. The broadband
emission of 3C~279 for this state indicates suppressed external
Compton emission, which makes it an ideal candidate for simulation
using a synchrotron-SSC model. We use physical jet parameters obtained
from the VLBA monitoring to guide our modeling efforts and discuss the
role of various intrinsic parameters and radiation processes in
producing the resultant SED.

We briefly describe the model of Joshi \& B\"ottcher (2011) in \S
\ref{model}. We discuss our findings about the connection between the
flux and polarization variation observed in 3C~279 during flaring and
quiescent periods of November 2008 and May 2010, respectively, in \S
\ref{3c2}. We discuss our first results from this study in \S
\ref{results}. We summarize and give a brief description of future
work in \S \ref{summary}.

\section{\label{model}Internal Shock Model}
\begin{figure}
\includegraphics[width=75mm]{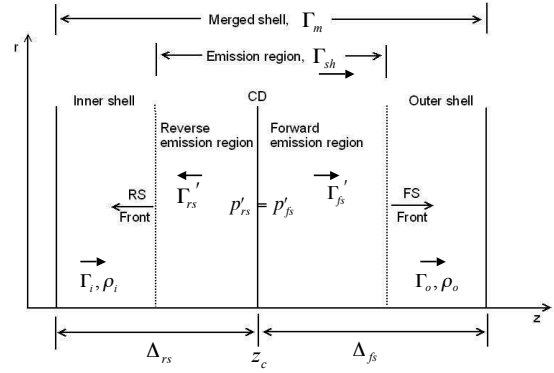}
\vspace{-10pt}
\caption{Schematic of the emission region with RS travelling into the
  inner shell and FS moving into the outer shell. The bulk Lorentz
  factors (BLFs) of the two shells are $\Gamma_{\rm i}$ \&
  $\Gamma_{\rm o}$, respectively, such that $\Gamma_{\rm i} >
  \Gamma_{\rm o}$. The primed quantities refer to the comoving frame
  and unprimed refer to the lab (AGN) frame. The comoving pressures
  $p_{\rm fs}^\prime$ and $p_{\rm rs}^\prime$ of the shocked fluids
  across the CD are considered equal. $\Delta_{\rm rs}$ and
  $\Delta_{\rm fs}$ are the widths of the inner and outer shells,
  respectively, after the collision. The quantities have been obtained
  by using hydro - dynamic and relativistic jump conditions across the
  shocks and CD \citep[]{sp2001, bd2010}.}
\vspace{-10pt}
\label{emission_region}
\end{figure}

The mode of acceleration of particles to highly relativistic energies
and its location in the jet is not yet completely understood. A
colliding shell or an internal shock model offers a way to gain
insight into the physics of particle acceleration. Under this model,
the central engine (black hole + accretion disk) is assumed to spew
out shells of plasma with different velocity, mass, and energy. An
inner shell, ejected at a later time, and travelling with a higher
speed catches up to a slower moving outer shell that is ejected at an
earlier time. The subsequent collision results in an emission region,
as shown in Figure \ref{emission_region}, with two internal shocks
(reverse (RS) and forward (FS)) separated by a contact discontinuity
(CD) and traveling in opposite directions to each other in the frame
of the shocked fluid. As the shocks propagate through the emission
region, they convert the ordered bulk kinetic energy of the plasma
into the magnetic field energy and random kinetic energy of the
particles. The highly accelerated particles then radiate and produce
the emission observed from the jet. The treatment of shell collision
and shock propagation is hydrodynamic and relativistic in nature
\citep[]{sp2001, bd2010}.

The model follows the evolution of the electron and photon populations
inside the emission region in a time-dependent manner. The equations
governing this evolution are,

\begin{eqnarray}
\label{2}
{\partial n_{e} (\gamma, t) \over \partial t} = -{\partial \over
  \partial \gamma} \left[\left({d\gamma \over dt}\right)_{loss} n_{e}
  (\gamma, t)\right] + Q_{e} (\gamma, t) - \nonumber\\
\frac{n_{e} (\gamma, t)}{t_{e,esc}}~
\end{eqnarray}
and
\begin{equation}
\label{3}
{\partial n_{ph} (\epsilon, t) \over \partial t} = \dot n_{ph,em}
(\epsilon, t) - \dot n_{ph,abs} (\epsilon, t) - \frac{n_{ph}
  (\epsilon, t)}{t_{ph,esc}}~.
\end{equation}

Here, $(d\gamma/dt)_{\rm loss}$ is the particle (referred to as
electrons in the rest of the text) energy loss rate due to synchrotron
and inverse Compton losses. The quantity $Q_{e} (\gamma, t)$ is the
sum of external injection of electrons and intrinsic $\gamma - \gamma$
pair production rate, and $t_{\rm e,esc}$ is the electron escape time
scale. The quantities $\dot n_{\rm ph,em} (\epsilon, t)$ and $\dot
n_{\rm ph,abs} (\epsilon, t)$ are the subsequent photon emission and
absorption rates, and $t_{\rm ph,esc}$ is the photon escape timescale
(see \S \ref{scheme}). The radiative energy loss rates and photon
emissivities are calculated using the time-dependent radiation
transfer code of Joshi \& B\"ottcher (2011).

The model reproduces the NIR to $\gamma$-ray emission from blazars as
it follows the evolution of the emission region out to sub-pc scales
only. Thus, it simulates the early phase of $\gamma$-ray
production. During this time, the radiative cooling is strongly
dominant over adiabatic cooling and the emission region is highly
optically thick at GHz radio frequencies. Hence, the simulated radio
flux is well below that of the actual radio data.

\subsection{\label{scheme}Multi-zone Radiation Transfer Method}
\begin{figure}
\includegraphics[width=80mm]{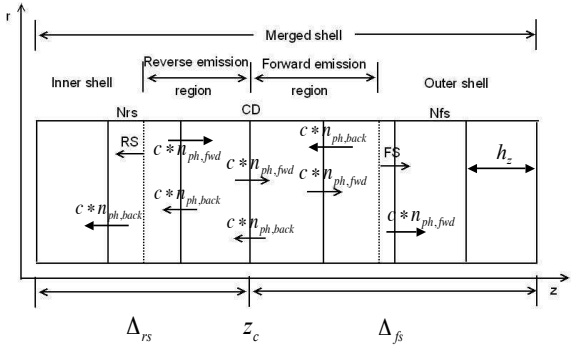}
\vspace{-10pt}
\caption{Schematic of the radiative transfer in between the zones
  using the appropriate photon escape probability function. The
  unprimed $n_{\rm ph, \sim fwd/back}$ values represent the photon
  densities in the forward and backward direction, respectively, in
  the comoving frame of the emission region. The rest of the unprimed
  quantities refer to the lab frame.}
\vspace{-10pt}
\label{zone_dia}
\end{figure}

We calculate the resultant spectrum from the collision, for a
cylindrical emission region, in a time-dependent manner. We consider
the inhomogeneity in the photon and electron energy density throughout
the emission region by dividing the region into multiple zones and
providing a fraction of the photon flux from each zone to its adjacent
zones in the two directions, as shown in Figure \ref{zone_dia}.

The fraction of photon flux from a zone or, in other words, the photon
escape rate of a zone in a particular direction is calculated using
the photon density of that zone and the probability of escape, P, for
a photon from that zone in that direction (forward, backward, or
sideways), according to the equation
\begin{equation}
\label{ndoteqn}
\frac{dn_{\rm ph, fwd/back/side}(\epsilon, \Omega)}{dt} = \frac{n_{\rm
    ph}(\epsilon, \Omega)}{t_{\rm ph, esc}} P_{\rm fwd/back/side}~,
\end{equation}
where $t_{\rm ph, esc}$ is the volume-and-angle-averaged photon escape
timescale for a cylindrical region that is calculated
semi-analytically in the model. We use this framework to calculate the
radiation transfer within each zone and in between the zones
\citep{jb2011}.

\section{\label{3c2}The Blazar 3C~279}
Analytical studies of the SEDs of 3C~279, collected during the
lifetime of CGRO, indicate that the SSC mechanism is responsible for
producing X-rays \citep[]{mcg1992, bm1996}. Intensive multiwavelength
campaigns \citep[see e.g.][]{lv2008, cr2008, co2007} and theoretical
efforts \citep[see e.g.][]{bp2009, ds1993} have provided insights on
the physical properties of 3C~279. But many aspects of the radiation
mechanism responsible for high-energy emission still remain unclear.
Also, no consistent trends of cross-correlations have been found
between optical, X-ray, and $\gamma$-rays \citep{hr2001} although the
source, in early 2009, exhibited coincidence of a $\gamma$-ray flare
with that in the optical and a dramatic change of optical polarization
angle during that time \citep{aa2010}. Such behavior poses a serious
challenge to the single-zone leptonic jet model approach where the
electron population is assumed to be homogeneous and is responsible
for producing synchrotron emission as well as high-energy emission
through inverse Compton scattering \citep{co2010}.

\begin{figure}
\includegraphics[width=75mm]{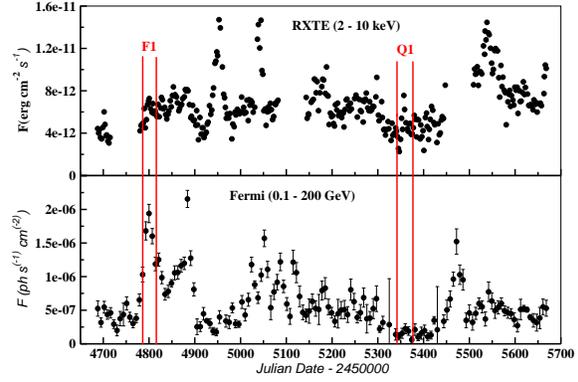}
\vspace{-10pt}
\caption{Observed lightcurves of 3C~279 from 08/09/2008 - 04/16/2011
  in $\gamma$- and X-ray energy bands. The top panel corresponds to
  the data compiled from RXTE while the bottom panel refers to the
  Fermi data. Long-term variability can be seen in the X-rays \&
  $\gamma$-rays.}
\vspace{-10pt}
\label{3c279lcs}
\end{figure}

\begin{figure}
\includegraphics[width=70mm]{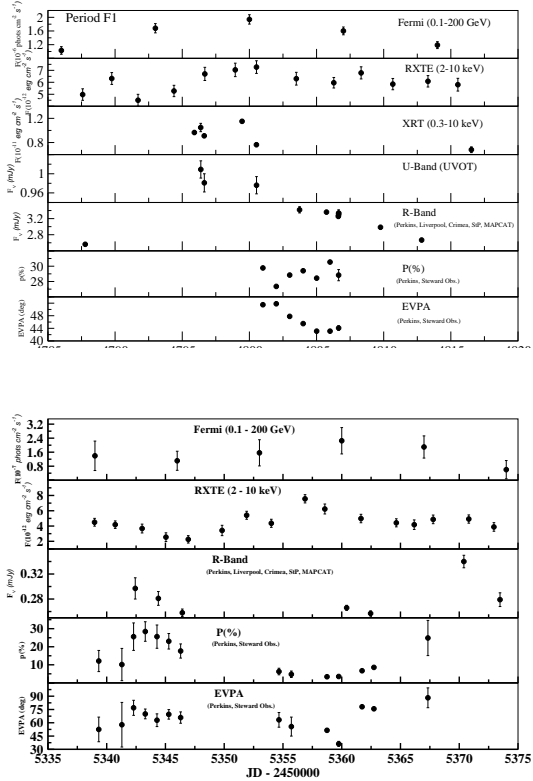}
\vspace{-10pt}
\caption{Multi-frequency lightcurves for the period Q1. R-band fluxes
  correlate with the optical polarization.}
\vspace{-10pt}
\label{q1lcs}
\end{figure}

In order to improve our insight into the behavior of 3C~279, we aim to
understand the evolution of the source at different flux levels by
simulating the corresponding SEDs, using a multi-zone leptonic jet
model, and looking for connections between the evolution of spectral
states and the behavior of the corresponding polarization at optical
and radio wavelengths. Figure \ref{3c279lcs} shows the longterm
$\gamma$- and X-ray lightcurves (LCs) of 3C~279 from August 2008 till
April 2011. From these LCs, we have extracted two 1-month time
periods: F1 (11/08/2008 - 12/08/2008) and Q1 (05/22/2010 -
06/26/2010), corresponding to a $\gamma$-ray flaring and a quiescent
period, as shown in Figures \ref{f1lcs} \& \ref{q1lcs}.

As can be seen from Fig. \ref{3c279lcs}, long-term variability exists
both in the X-rays \& $\gamma$-rays. On the other hand, in case of F1
(see Fig. \ref{f1lcs}, more complicated features exists for such
shorter time-periods that need to be considered in detail. While no
correlation is apparent between the flux and polarization behavior at
optical and radio wavelengths for period F1, the VLBA data on
06/14/2010 (period Q1) resulted in a degree of polarization $p = 4.7
\pm 0.8 \%$ and a radio polarization angle $EVPA = 78.3 \pm 4.9^{o}$,
which agrees very well with the corresponding values of \textit{p}
($8.6 \pm 0.3 \%$) and \textit{EVPA} ($75.8 \pm 0.9^{o}$) at optical R
band. This suggests a common source of origin for the two.

\begin{figure}
\vspace{-30pt}
\includegraphics[width=70mm]{f5}
\vspace{-10pt}
\caption{Multi-frequency lightcurves for the period F1.}
\vspace{-10pt}
\label{f1lcs}
\end{figure}

\begin{figure}
\includegraphics[width=75mm]{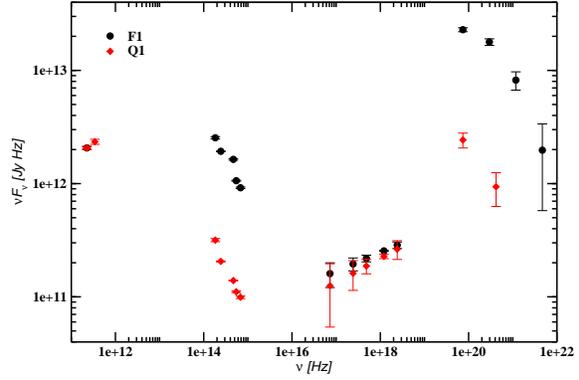}
\vspace{-10pt}
\caption{The observed SEDs of 3C~279 corresponding to periods F1 and
  Q1.}
\vspace{-10pt}
\label{vlbaim}
\end{figure}

The SEDs of 3C~279 corresponding to states F1 and Q1 are shown in
Fig. \ref{vlbaim}. The data have been compiled from observations with
Fermi, Swift, VLBA, and various ground-based optical and radio
telescopes. The optical and near-IR data were dereddened using the
Galactic extinction coefficients of Schlegel et al. (1998) and the
magnitudes were converted to fluxes using the zero-point
normalizations of Bessel et al. (1998). The shape of the optical and
near-IR spectra indicates that the synchrotron component for both
states peaks in the near-IR regime with the $\nu F_{\nu}$ value of F1
higher than that of Q1 by almost an order of magnitude. The X-ray flux
of F1 remains in a low state coinciding with that of state Q1. The
shape of the optical spectrum suggests that the synchrotron component
extends into the UV regime with a turnover to the high-energy
component taking place at $\sim 10^{16}$ Hz. The $\gamma$-ray spectrum
for both states indicates a higher total energy output in the
high-energy component compared to the low-energy one, which is typical
for the class of flat spectrum radio quasars \cite{bm2007}.

\subsection{\label{results}First Results}
The blazar 3C~279 was observed in an extensive multi-frequency
campaign in January 2006 \citep{co2010}. The object went into an
optical flaring state on 01/15/2006. The interesting feature of this
state is that the X-ray flux remained in a low state during that
period, while coinciding with the historical X-ray low-state of June
2003 \cite{co2010}. Almost a month later, in Februrary 2006,
it was detected by Major Atmospheric Gamma-ray Imaging Cherenkov
Telescope (MAGIC) in a very high energy (VHE) $\gamma$-ray state
\citep{al2008}.

\begin{figure}
\includegraphics[width=75mm]{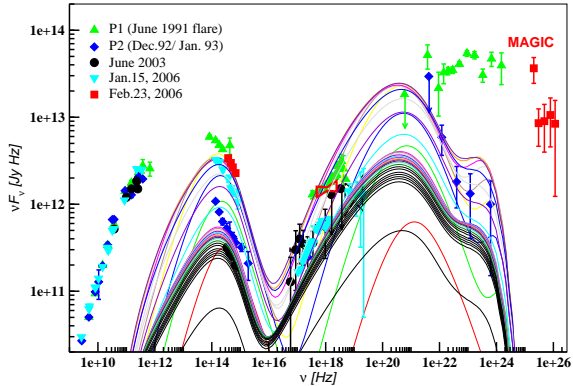}
\vspace{-10pt}
\caption{Simulated instantaneous SEDs of 3C~279 for 01/15/2006 showing
  the time-dependent evolution of the optical high state. Each solid
  curve is representative of the spectrum corresponding to a
  particular instant of time.}
\vspace{-10pt}
\label{3c279inst}
\end{figure}

\begin{figure}
\includegraphics[width=75mm]{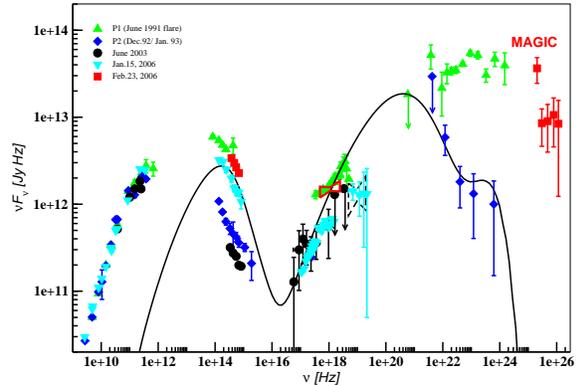}
\vspace{-10pt}
\caption{Simulated time-integrated SED, averaged over 1 day, of 3C~279
  for 01/15/2006.}
\vspace{-10pt}
\label{3c279int}
\end{figure}

As a first step toward simulating the SEDs of 3C~279 in order to
understand the evolution of spectral states and compare the flaring
states of the source in November 2008 and January 2006, we have used
the time-dependent leptonic jet model of Joshi \& B\"ottcher (2011)
that calculates the synchrotron and SSC radiation to reproduce the
observed SED of 3C~279 corresponding to the optical high state of
01/15/2006. Figures \ref{3c279inst} \& \ref{3c279int} show the
instantaneous and time-integrated simulated SEDs of 3C~279,
respectively, for this day. The model independent parameters
\citep{bh2005} estimated using the SED, VLBA observations
\citep{js2005}, and variability on 1-day timescale were used to
develop an initial set of input parameters:

\begin{eqnarray}
\delta &\approx& 15.5 \hss \cr
R &\approx& 2.5 \times 10^{16} \; {\rm cm} \hss \cr
B &\approx& 0.82 \, \epsilon_B^{2/7} \; {\rm G} \hss \cr
\gamma_{\rm min} &\approx& 9.0 \times 10^2 \hss \cr
\gamma_{\rm max} &\approx& 2.2 \times 10^4 \hss \cr
p &\approx& 4.3 \hss \cr
\theta_{\rm obs} &\approx& 2.1^{o} \pm 1.1^{o}~. \hss
\label{param_sum3c}
\end{eqnarray}

Here, $\delta$ is the Doppler boosting factor. The symbol R is the
radius of the emission region, B is the comoving magnetic field of the
region, and $\epsilon_{\rm B}$ is the equipartition parameter between
the magnetic field and the electron energy density, assumed to be
equal to 1 here. The quantities $\gamma_{\rm min}$ \& $\gamma_{\rm
  max}$ refer to the low and high energy cutoffs of the electron
energy distribution. The spectral index of the electron population, at
the time of injection, is given by \textit{q} and \textit{p} is the
equilibrium spectral index obtained using the optical synchrotron
spectrum, which is given by $p = q + 1$ for strongly cooled
electrons. The quantity, $\theta_{\rm obs}$ is the observing angle
inferred from VLBA observations.

The entire emission region has been divided into 100 zones (50 in the
forward and 50 in the reverse emission region) to analyze the observed
SED of the source. The initial set of parameters was modified to
reproduce the state of 3C~279 as observed on 01/15/2006. Various
simulations were carried out by varying input parameters one at a time
to obtain a reasonable fit to the observed SED of 3C~279. A fit is
considered successful if it passes through the observed data points,
corresponding to 01/15/2006, without over-producing the X-ray flux. An
observer's line of sight angle, $\theta_{\rm obs} = 1.6^o$, and a
cross-sectional radius, $R = 6.85 \times 10^{16}$~cm, for the
cylindrical emission region, have been chosen to obtain the current
best fit. Table~\ref{paramlist} lists the rest of the parameters used
for obtaining the current fit of the observed SED. These parameters
result in $\Gamma_{\rm sh} \approx 18.6$ for the entire emission
region, $B \approx 1.3 ~G$ and $\gamma_{\rm max} \approx 1.2 \times
10^5$ for both the forward and reverse emission regions, and
$\gamma_{\rm min, ~fs} \approx 1.4 \times 10^3$ \& $\gamma_{\rm min,
  ~rs} \approx 1.7 \times 10^3$ for the forward and reverse emission
regions, respectively.

As can be seen from Figures \ref{3c279inst} \& \ref{3c279int}, the
observed SED for 01/15/2006 shows a high-energy bump that is
indicative of a dominant SSC component and a suppressed EC
component. The time-integrated simulated SED of the successful model
passes very close to the IR and optical data points, indicating that
the synchrotron component is responsible for the lower energy bump of
the SED. The spectral upturn takes place in the soft X-rays at $\geq$
0.1 keV due to the presence of the SSC component in the
simulation. The lower-energy part of the SSC component reproduces the
X-ray data quite well, suggesting the dominance of the SSC component
in producing this part of the high-energy bump. As can be seen in
Figure \ref{3c279int}, there is a break in the low-energy part of the
X-ray spectrum of January 2006 at $\sim 10^{19}$ Hz, which could be
due to the presence of a low-energy cutoff of the electron
distribution at ultrarelativistic energies \citep{co2010}.

\begin{table}
\begin{center}
\caption{Model parameters used to reproduce the state of 3C~279
  as observed on 01/15/2006.\\}
\begin{tabular}{|l|c|c|c|c|c|c|c|}
\hline \textbf{Source} & \textbf{$L_{w}$} & \textbf{$\Gamma_{i}$} & \textbf{$\Gamma_{o}$} & \textbf{q} & \textbf{$\varepsilon_{e}$} & \textbf{$\varepsilon_{B}$} & \textbf{$\zeta_{e}$}
\\
\hline & [$10^{48}$~ergs/s] & & &  & [$10^{-1}$] & [$10^{-4}$] & [$10^{-2}$]\\
\hline 3C279  &  1  &  38   &  10   &  4.6    &  4   &  5   &  8\\
\hline
\end{tabular}
\\
  $L_{w}$: luminosity of the injected electron population in the blob,
  $\Gamma_{i,o}$: BLFs of the inner and outer shells before collision,
  q: particle spectral index, $\varepsilon_{e}$: ratio of electron and
  shock energy density, $\varepsilon_{B}$: ratio of magnetic field and
  shock energy density, $\zeta_{e}$: fraction of accelerated
  electrons 
\label{paramlist}
\end{center}
\end{table}

\section{\label{summary}Discussion and Future Work}
We have presented first results of our study on the behavior of 3C~279
at different spectral states, and the connection between the flaring
and quiescent states and the behavior of its polarization at optical
and radio wavelengths.

In order to carry this out, we have extracted a $\gamma$-ray flaring
(F1) and a quiescent (Q1) state from the long-term LCs of 3C~279 shown
in Fig. \ref{3c279lcs}. In case of F1, a clear cross-correlation
cannot be seen at optical frequencies over a shorter time-period such
as a month (see Fig. \ref{f1lcs}) and some complicated features exist
that need to be considered in detail. In case of Q1, there is an
anti-correlation between $\gamma$- and X-ray, with X-rays mildly
flaring during the $\gamma$-ray quiescent state, but the R-band flux
shows a strong relationship to the optical polarization (see
Fig. \ref{q1lcs}). On the other hand, while no correlation can be seen
between the flux and polarization behavior at optical and radio
wavelengths for the period F1, the VLBA data during the Q1 period
agrees very well with the corresponding values of \textit{p} and
\textit{EVPA} at optical R band. This suggests a common source of
origin for the two during this state.

In order to understand the behavior of the source at different
spectral states, it is necessary to simulate the corresponding SEDs of
the source. To that end, we have simulated the optical high state of
3C~279 observed in January 2006 using the time-dependent multi-zone
leptonic jet model of \cite{jb2011}. The model, in its current form,
calculates radiation resulting from synchrotron and SSC processes. The
spectral state of 3C~279 observed in January 2006 makes a good
candidate for this model, as the $\gamma$-ray emission is suppressed
in this case and the X-ray emission is generally associated with SSC
scattering \citep[]{mcg1992, bm1996, co2010}.

Our current model of the SED results in a high value of $L_{\rm w}$
compared with that of Collmar et al. (2010). This is reasonable, as
the internal shock model is expected to have low acceleration
efficiency, and thus would need a higher kinetic luminosity to achieve
the desired acceleration. The current fit results in a value of
magnetic field that agrees reasonably well with the model independent
parameter estimate and that of Collmar et al. (2010). The value of
electron energy index, \textit{q}, used to obtain the current fit is
almost equal to the model independent parameter \textit{p} and is
comparatively higher than that of Collmar et al. (2010). This
indicates that the synchrtron spectrum in the current fit does not
come from a cooled population of electrons. The value of \textit{q}
needs further adjustment in order to obtain a satisfactory fit. With a
successful fit, we hope to gain insights on the acceleration
efficiency, the relative speeds with which the two shocks are moving
away from each other (in the frame of the shocked fluid), and the mass
of each plasma shell needed to reproduce an optical high state of
3C~279 such as that of 01/15/2006.

As part of our future work, the external Compton component due to
photons entering the jet from the BLR and dusty torus is being
incorporated in the existing model of \cite{jb2011} to reproduce more
fully the SEDs of blazars, especially for flat spectrum radio quasars
(Joshi et al. 2012, in prep.), and to analyze a wider variety of
blazar spectral states. We plan to use the extended version of the
model to simulate the SEDs of 3C~279 for a typical period of \~ 10
days corresponding to F1 and Q1 and compare the results with that of
the January 2006 optical high state of 3C~279.

Further, we plan to incorporate the effects of magnetic field
orientation, as inferred from polarization monitoring programs, on the
resultant spectral variability and SEDs of blazars. This would aid us
in the study of intrinsic parameter differences between various blazar
subclasses that arise from the orientation of the magnetic field in
the jet.

\bigskip 
\begin{acknowledgments}
This research was supported in part by NASA through Fermi grants
NNX10AO59G, NNX08AV65G, and NNX08AV61G and ADP grant NNX08AJ64G, and
by NSF grant AST-0907893.

\end{acknowledgments}

\bigskip 

\end{document}